\documentstyle[preprint,aps]{revtex}
\begin{document} 
\tightenlines
\draft
\title{Event anisotropy in 4.2A GeV/c C+C collisions}
\author{ Lj. Simi\'c 
 and J. Milo\v{s}evi\'c }
\address{Institute of Physics, P.O. Box 68, 11080 Belgrade, Yugoslavia}
\maketitle
\begin{abstract} 
 
 The directed and elliptic flow of protons and negative pions 
 in 4.2A GeV/c  C+C collisions is studied  using the Fourier 
 analysis of azimuthal distributions. It is found that the protons exhibit  
 pronounced directed flow, while the flow of pions is 
 either non existent or too weak to be detected 
 experimentally. 
 Also, it is found that in the entire rapidity interval the elliptic
 flow is very small if not zero. These results are confirmed 
 by the Quark-Gluon-String Model (QGSM) 
 and the relativistic transport model  (ART 1.0),  
 except that these models predict very
 weak antiflow of pions. The more detailed comparison with the QGSM 
 suggests that the decay of 
 resonances  and rescattering of secondaries dominantly determine  
 the proton and negative pion flow at this energy.

\end{abstract}
\pacs{PACS number: 25.75.Ld}

\clearpage

\indent
      Event anisotropy, often called flow, has been observed in heavy-ion 
  collisions
  at all incident energies \cite{Part,Chance,Reis,Barr,Appel,Aggar}. At 
  Bevalac energies and
  below, the flow is usually studied in terms of the mean in-plane component of
  transverse momentum at a given rapidity, $\langle p^x(y) \rangle$,
  \cite{DanOdy} and additionally quantified in terms of derivative at 
  the midrapidity
  $F_y=d\langle p^x \rangle/dy$. At high energies, the Fourier
  expansion of the azimuthal distribution of particles is used
  \cite{Ollit92,Zhang,PosVol}. In this expansion the first harmonic, $v_1$, 
  quantifies
  the directed flow while the second harmonic, $v_2$, quantifies the 
  elliptic flow.
  Using the Fourier expansion, the anisotropic transverse flow was analysed for
  heavy symmetric systems at the AGS \cite{Barr} and SPS \cite{Appel,Aggar} 
  energies. It was found that this anisotropy, and particularly the elliptic 
  flow, plays an important
  role for investigating properties of hadronic matter
  \cite{Pinken,Dan98,Li,Liu}.
  However, it is 
  still not clear whether the experimentally observed event anisotropy
  is of a dynamic origin or is due to the shadowing of spectator matter, 
  passing time, etc. 
 
        In this paper the directed and elliptic flow of protons and 
  negative pions 
  in 4.2A GeV/c C+C collisions is studied  using the Fourier 
  analysis of azimuthal distributions. The analysis is performed using 9500 
  C+C semicentral and central collisions 
  obtained with the 2-m propane
  bubble chamber, exposed at JINR, Dubna synchrophasotron.
  The data for semicentral and central collisions roughly correspond 
  to the upper $50\%$  of the inelastic cross-section.
  Additionally, the same type of analysis is performed using the 400000 events 
  generated by the 
  QGSM \cite{AGT,AGST,Amel} and 200000 events generated by the 
  relativistic transport model (ART 1.0) \cite{Bao}. 
  For these events the same centrality criterion is applied as in experiment,
  leading to the average impact parameter $\approx$ 2.6. 
  In the 2-m propane bubble chamber practically all 
  charged reaction products are detected. Negative particles, 
  except identified electrons
  are considered to be $\pi^-$. Among them remains admixture of 
  unidentified fast
  electrons ($<$ 5\%). All positive particles with momenta less than 0.5 GeV/c
  are classified either as protons or $\pi^+$ mesons according to their
  ionisation density and range. 
  Positive particles above 0.5 GeV/c are taken to
  be protons, and because of this, the admixture of $\pi^+$ of 
  approximately 18\%  is subtracted statistically using the
  $\pi^+$ and $\pi^-$ momentum distributions. From the resulting number 
  of protons,
  the projectile spectators (protons with momenta $p>$ 3 GeV/c and emission
  angle $\theta<4^{\rm o}$) and target spectators (protons with momenta
  $p<$ 0.3 GeV/c) are further subtracted. The resulting number of participant
  protons still contains some 4\%  of deuterons 
  (with $p>$ 0.48 GeV/c) which are statistically subtracted. The
  admixture of tritons (with $p>$ 0.65 GeV/c) is not considered. 
  The experimental
  data are also corrected to the loss of particles emitted at small angles
  relative to the optical axes of chamber. The aim of this correction is 
  to obtain isotropic
  distribution in azimuthal angle and smooth distribution in emission
  angle (both measured with respect to the direction of the incoming
  projectile).
 
    The azimuthal distribution of particles may be represented with the 
 first three 
 terms of the corresponding Fourier expansion
 \begin{equation}
 \label{Fourexp}
 \frac{dN}{d\phi}\approx\frac1{2\pi}\big [1+2v_1\cos(\phi)+2v_2\cos(2\phi)
 \big ],
 \end{equation}
 where the two coefficients, $v_1$ and $v_2$, quantify 
 the directed 
 and elliptic flow via $v_1=\langle cos(\phi) \rangle$ and 
 $v_2=\langle cos(2\phi) \rangle$. In Eq. (1), 
 $\phi=\phi_{lab}-\Phi_{plane}$ is the particle azimuthal angle determined
 with respect to the reaction plane, with $\phi_{lab}$ denoting the azimuthal 
 angle of particle in the laboratory frame and $\Phi_{plane}$ 
 denoting the azimuthal angle of the (true) reaction plane.
 Since both the projectile momentum and the 
 impact parameter vectors are 
 available in the QGSM simulation, they are used to determine the corresponding
 reaction plane. 
 In the experiment the reaction plane is determined, for each event, 
 using the projectile momentum vector and the
 vector $\bf Q$ determined from \cite{DanOdy}
 \begin{equation}
 \label{vecQ}
 {\bf Q}=\sum_i {\bf p}_{Ti} (y>y_{cm}+\delta)-\sum_j{\bf p}_{Tj} 
 (y<y_{cm}-\delta),
 \end{equation}
 where ${\bf p}_{T}$ represents the transverse momentum of the proton 
 emitted in the forward ($y>y_{cm}+\delta$), or backward ($y<y_{cm}-\delta$),
 hemisphere.  Here, $y_{cm}$ denotes the center of mass rapidity 
 while the quantity  $\delta$ (=0.2)  removes the protons 
 emitted around the $y_{cm}$ which are not 
 contributing to the determination of the reaction plane.  
 The reaction plane angle for a proton is determined using this expression  
 only if this proton is not included in the above sum (i.e. if its rapidity 
 lies in the interval from $y_{cm}-\delta$ to $y_{cm}+\delta$). Otherwise, 
 in order to avoid autocorrelation (which is an effect of the finite 
 multiplicity), the 
 $\bf Q$ vector is constructed by the analogous expression in which the 
 contribution of this proton is simply omitted \cite{DanOdy}. 
 We found that the reaction plane angle distribution is essentially flat, 
 thus confirming the absence of significant distorsions which could 
 influence the magnitude of the extracted flow parameters. 
    
      The accuracy with which  the reaction plane angle is  determined, 
 i.e. the reaction plane resolution, is evaluated
 by the subevent method \cite{DanOdy}. 
 In this method, each event is divided randomly into two subevents, and 
 then the corresponding two reaction planes are determined. Subsequently, 
 the absolute value of the
 relative azimuthal angle, $\Phi_{12}$, between these two estimated reaction
 planes is obtained. The width, $\sigma$, of the $\Phi_{12}$ 
 distribution determines the reaction plane resolution.
 For C+C  collisions we find $\sigma=50^0$.  
 The relative azimuthal angle distribution is the basis for 
 the correction of the Fourier coefficients, $v'_n$, 
 obtained with the estimated reaction plane.
 The relationship between the $v'_n$, and the Fourier
 coefficients $v_n$ obtained relative to the true reaction plane, is
 $v'_n=v_n\ \langle cos(n\Delta\Phi)\rangle$, 
 where  $\langle cos(n\Delta\Phi)\rangle$ is the correction factor 
 determined from $\Phi_{12}$ distribution following the prescription 
 given in \cite{PosVol,Ollit}. We find $\langle cos(\Delta\Phi)\rangle$=0.56 
 and $\langle cos(2\Delta\Phi)\rangle$=0.24.
 The correctness of this procedure is checked using the QGSM. 
 Using this model, the coefficients $v_1$ and $v_2$ are calculated
 with respect to the true reaction plane and also with respect to the estimated
 reaction plane. The results of the comparison will be discussed below. 

       Fig. 1 (top) displays the experimentally determined $v_1$  
 coefficient
 vs. $y$ (with $y$ calculated in the
 center-of-mass frame), for protons and negative pions
 together with the $v_1$ calculated 
 with QGSM relative to the true reaction plane and relative to the 
 estimated reaction plane. 
 For the proper comparison with the experiment, we excluded protons  
 satisfying  cuts for the proton spectators in the experiment. 
 In the case of protons
 it is seen that the values of the two QGSM results for $v_1$  
 are quite close. 
 The dependence of $v_1$ on rapidity is characterised by a curve
 with a positive slope and with 
 the zero-crossing at $y=0$. The curve indicates a positive directed 
 flow with magnitude $v_1\approx 0.17$, at rapidities close to 
 the beam rapidity ($0.7<y<1.5$).  
 The QGSM reproduces satisfactorily the shape 
 of $v_1(y)$ curve and the magnitude of the flow.
 Using the extracted values of $v_1$ and their relation
 to the mean transverse momentum projected onto the reaction plane,  
 $v_1=\langle p_x \rangle /\langle p_T \rangle$, we can evaluate 
 $\langle p_x\rangle$ as a 
 function of rapidity and determine the slope,  
 $F=d \langle p_x \rangle /d(y/y_b)$, 
 with respect to rapidity normalised to beam rapidity in the lab frame 
 ($y_b$=2.2). 
 In the present analysis we find for the slope at the midrapidity 
 $F=$ 144 MeV/c.
 After the normalisation to the mass number of the colliding system 
 we obtain the so called scaled flow $F_S = F/(A_1^{1/3} +A_2^{1/3})=$
 31 MeV/c.  This value is in agreement with 
 the observed trend \cite{Chance} that after reaching the maximum
 at beam energy 
 around 0.7-2A GeV, the directed flow slowly decreases with increasing beam 
 energy. 
 
 For negative pions the experimental values of $v_1$ indicate that 
 the directed flow is non-existent. 
 This result is confirmed by the model calculations of $v_1$ with respect to
 the estimated reaction plane. However, the model calculations of $v_1$ 
 with respect to the true reaction plane show the existence of a very weak 
 directed 
 flow of pions with the sign of $v_1$ opposite to that of protons and with
 the maximum value of 0.02 around the target rapidity. 
 This further suggests that in the collisions of light nuclei, like C+C, 
 the very weak flow, if it exists, is not measurable because of the 
 limited accuracy in determination of the reaction plane. 
 
      Fig. 1 (bottom) displays the experimentally determined $v'_2$ 
 coefficient
 vs. $y$  for protons and negative pions. This coefficient is 
 not corrected to the reaction plane resolution since the comparison
 of the model calculation of $v'_2$, obtained as in the   
 experiment, and the model calculation of $v_2$, relative 
 to the true reaction plane, 
 indicates that the corresponding correction 
 procedure for $v_2$, as outlined above, is not applicable. 
 The reason for this is the lightness 
 of the colliding nuclei and the smallness of the elliptic flow.
 The uncorrected values of $v'_2$ show  that  in the entire rapidity interval 
 the elliptic 
 flow is small ($|v^{'}_{2}|\leq 0.02$) if not zero, and this is 
 consistent with 
 the predictions of QGSM. 
 The experimental values for $v'_2$ 
 also show that, for both protons and pions, the elliptic flow 
 depends on rapidity and that around the beam and target rapidities 
 it is positive for protons and negative for pions.
 The positive sign for protons indicates an enhanced emission in the 
 reaction plane, while the negative sign for pions indicates 
 an enhanced emission perpendicular to the reaction plane.
 This behaviour points out to the shadowing by the nuclear matter 
 as the origin of the elliptic flow.

     Since the QGSM predictions are in fair agreement with the 
 experimental results at 4.2A GeV/c, we use 
 this model to clarify the question which of the  processes are responsible 
 for the flow effect. In this model, in collisions of light C+C 
 nuclei, approximately
 $40\%$ of protons and 
 $\approx 70\%$ of $\pi^-$  originate from 
 decay of the lowest-lying
 resonances ($\Delta'$s, $\varrho,\omega,\eta$ and $\eta^{'}$). 
 The rest originates from the 'non-resonant' primary and secondary 
 interactions of the type: 
 $NN\rightarrow NN\pi$, $\Delta N\rightarrow \Delta N$,  
 $\pi N \rightarrow \pi N$, $\pi NN\rightarrow NN$.
 The protons and pions from primary interaction  
 escape the collision zone without further rescattering and comprise 
 $\approx 5 \%$  of the total.
 Therefore, according to QGSM, we separately evaluate the flow of protons and 
 pions 
 originating from the following  sources:({\it i}) decay of resonances,
 ({\it  ii }) primary non-resonant interactions
 ({\it iii }) and secondary non-resonant interactions.

 Figure. 2 (top) shows  $v_1$ vs. rapidity  for  
 protons and negative pions originating from decay of resonances, and from 
 primary and secondary non-resonant interactions, as well as the overall
 $v_1$ for protons and $\pi^-$. 
 (In these model calculations, the experimental cuts for the proton 
 spectators were not applied, and this leads to a small difference 
 between the two curves for overall $v_1$ for protons in Figs. 1 and 2). 
 The protons originating both from the decay of resonances and  
 from the secondary interactions
 show the directed flow of similar intensity. The same applies to the 
 antiflow of pions.
 The protons from the primary interactions show a relatively flat 
 $v_1(y)$ distribution,  
 while the pions from these interactions show a strong directed antiflow  
 with magnitude $\approx -0.13$ around beam rapidity.  
 The antiflow of pions
 can be also explained by the shadowing effect, but the shadowing 
 matter is different  for pions from
 primary and secondary interactions since these pions are produced at 
 different 
 collision stages. Initially 
 (at times less than the passing time, $t_p$=4.2 fm/c) 
 the pions  are  shadowed by the cold spectators. 
 Later, after the spectator matter leaves the collision zone, 
 the pions are shadowed by the participant nucleons.
 This may be the underlying mechanism that leads to the different behaviour
 of $v_1$ for pions. 
 In heavy nuclei collisions, additionally generated by the QGSM,  
 the protons from primary interactions, similarly to the case of pions 
 from primary interactions, show strong antiflow caused by the shadowing 
 from the cold nuclear matter. 
 In the collisions of light nuclei this shadowing is small 
 and there is no preferential emission of the protons.
 
   Figure.2 (bottom) shows  $v_2$  vs. rapidity for  
 protons and negative pions originating from decay of resonances, and from 
 primary and secondary non-resonant interactions, as well as the overall
 $v_1$ for protons and $\pi^-$.
 It is seen that the particles
 from secondary interactions and from decay of resonaces,  
 exhibit similar behaviour.  
 The particles from the primary interactions show a clear  negative 
 elliptic flow, and 
 this out-of-plane emission can be attributed to the  
 shadowing by the cold spectators.
 
 In order to establish a less model dependent picture, the results of 
 the experiment are also compared with the relativistic transport model,
 ART 1.0. These are shown in Fig. 3, where the calculations are performed 
 both in the cascade  and in the so-called 'mean field' mode. The cascade mode 
 underestimates the magnitude of the proton flow (some 20$\%$), 
 and predicts a  
 small directed antiflow of negative pions with magnitude $v_1\approx 0.04$. 
 In the mean 
 field mode the model increases the magnitude of the proton flow. 
 Also, the ART model predicts a very small
 proton and pion elliptic flow ($|v_{2}|\leq 0.01$).

     In summary, the directed and elliptic flow of protons and negative pions 
  in 4.2A GeV/c C+C collisions was examined using the 
  Fourier analysis of azimuthal distributions of experimental events,  
  and also by using the events generated by the QGSM and ART 1.0 model. 
  It was found that the protons exhibit 
  strong directed flow with  magnitude  $v_1\approx 0.17$ at rapidities 
  close to the beam rapidity.
  The QGSM reproduces satisfactorily the shape 
  of the $v_1(y)$ curve and the magnitude of the flow. The ART  
  model underestimates this magnitude in the cascade mode and 
   increases this magnitude in the mean field mode.    
  For negative pions the flow  is 
  either non existent, or too weak to be detected 
  experimentally  due to the limited accuracy in the 
  determination of the reaction plane. The latter was suggested by the QGSM, 
  where calculations with respect to the estimated reaction 
  plane predicted non existant flow, while the calculations 
  with respect
  to the true reaction plane predicted
  a  small directed antiflow with magnitude $v_1\approx 0.02$.
  The predictions of the ART model are similar.
  Also, it was found that in the entire rapidity interval the elliptic 
  flow is small ($|v^{'}_{2}|\leq 0.02$) if not zero, and this is in 
  agreement with 
  the predictions of QGSM and ART 1.0. 
  According to the QGSM, the two factors that dominantly determine the 
  proton and negative pion flow, at this energy, 
  are the decay of resonances and the rescattering of secondaries. 
  The shadowing by the cold spectator matter affects 
  only the flow of the particles produced at the early stage 
  of the collision.\\

      The authors are grateful to members of the JINR Dubna group
  that participated in data processing, G. \v Skoro for making the QGSM code
  available to us, and I. Menda\v s for useful comments.

 \clearpage

 \begin{figure}
 \caption{ Rapidity dependence of $v_1$ and $v_2$ for protons and 
 $\pi^-$ for 4.2A GeV/c C+C collisions:
 $top$- filled circles represent the experimental results for $v_1$
 while the solid (dashed) line represents the    
 QGSM calculation for $v_1$ with respect to the true (estimated) 
 reaction plane;
 $bottom$- filled circles represent uncorrected experimental $v^{'}_2$ values 
 (see text), while the  
 solid (dashed) line represents the QGSM calculation for $v_2$ ($v^{'}_2$) 
 with respect to the true (estimated) reaction plane. 
 }   
 \label{fig1}
 \end{figure}

 \begin{figure}
 \caption{ Rapidity dependence of $v_1$ and $v_2$ for protons and 
 $\pi^-$ ({\it solid line}); for protons and 
 $\pi^-$  originating from decay of resonances ({\it stars}),
 primary non-resonant interactions ({\it full circles}),
 and secondary
 non-resonant interactions ({\it open circles}),
 for 4.2A GeV/c C+C collisions generated 
 with the QGSM. 
 }   
 \label{fig2}
 \end{figure}

 \begin{figure}
 \caption{ Experimental results for $v_1$, $v_2$ as a function of rapidity, 
  compared with ART 1.0 model calculations.
 }
 \label{fig3}
 \end{figure}

 
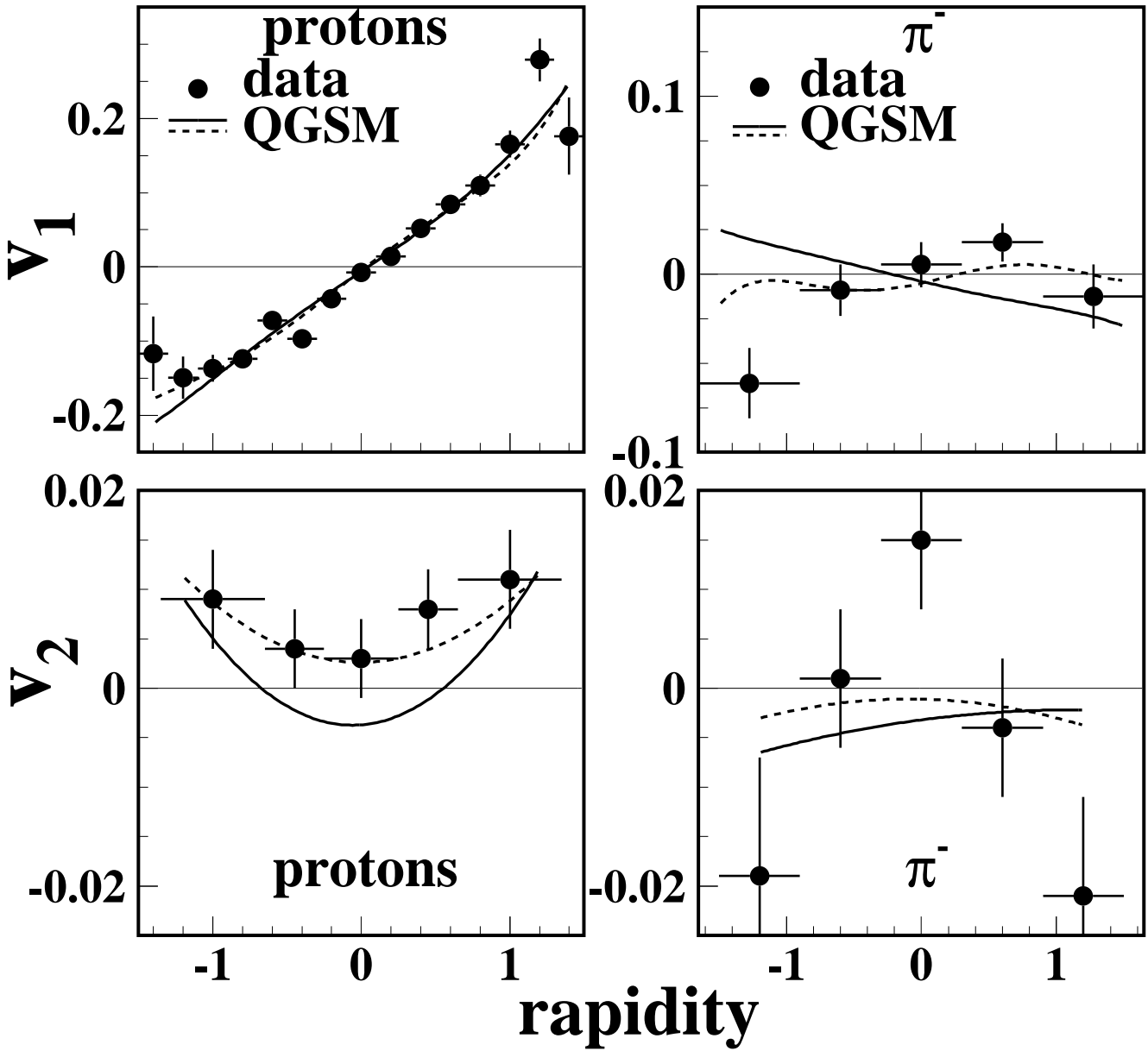
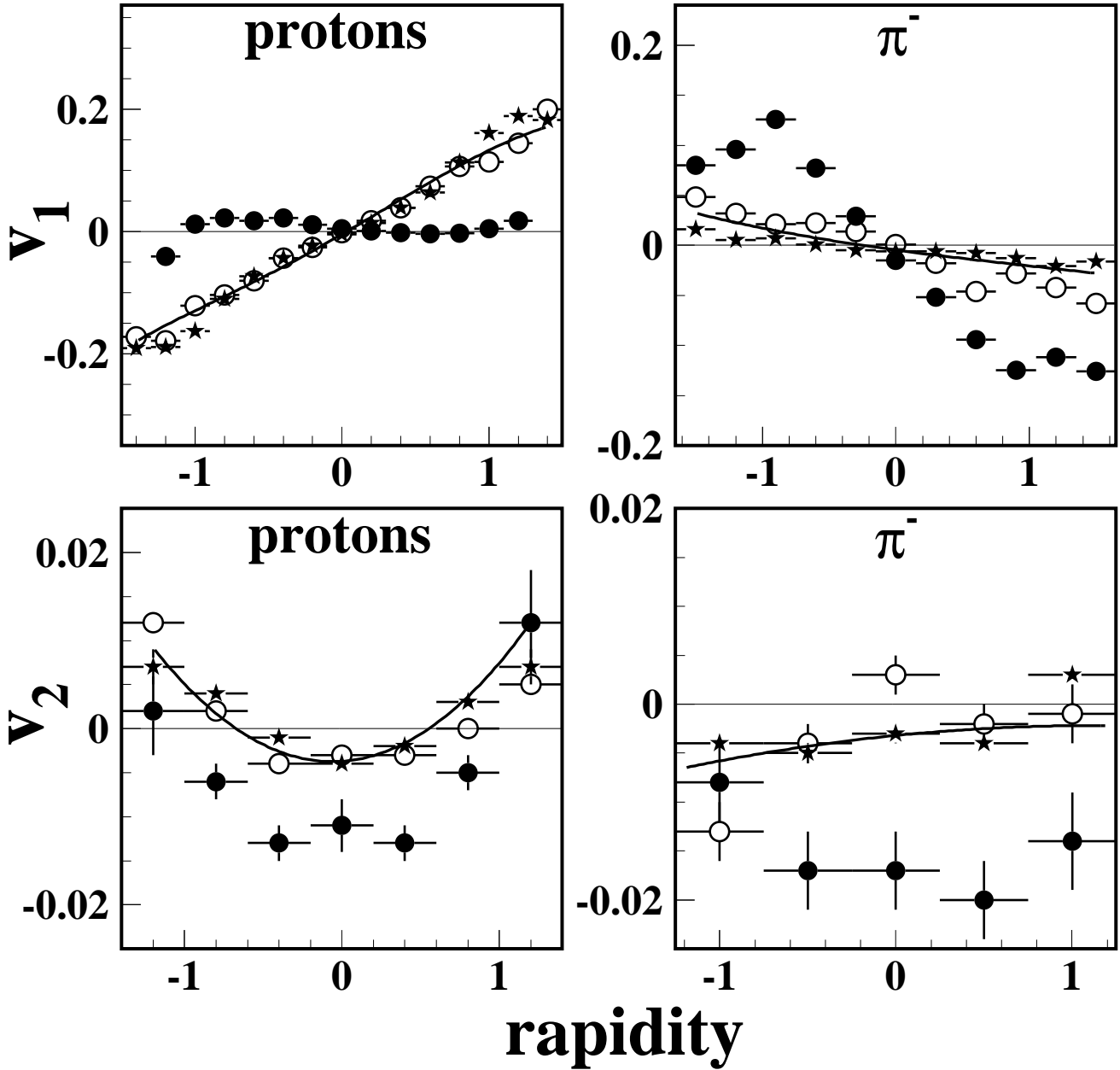
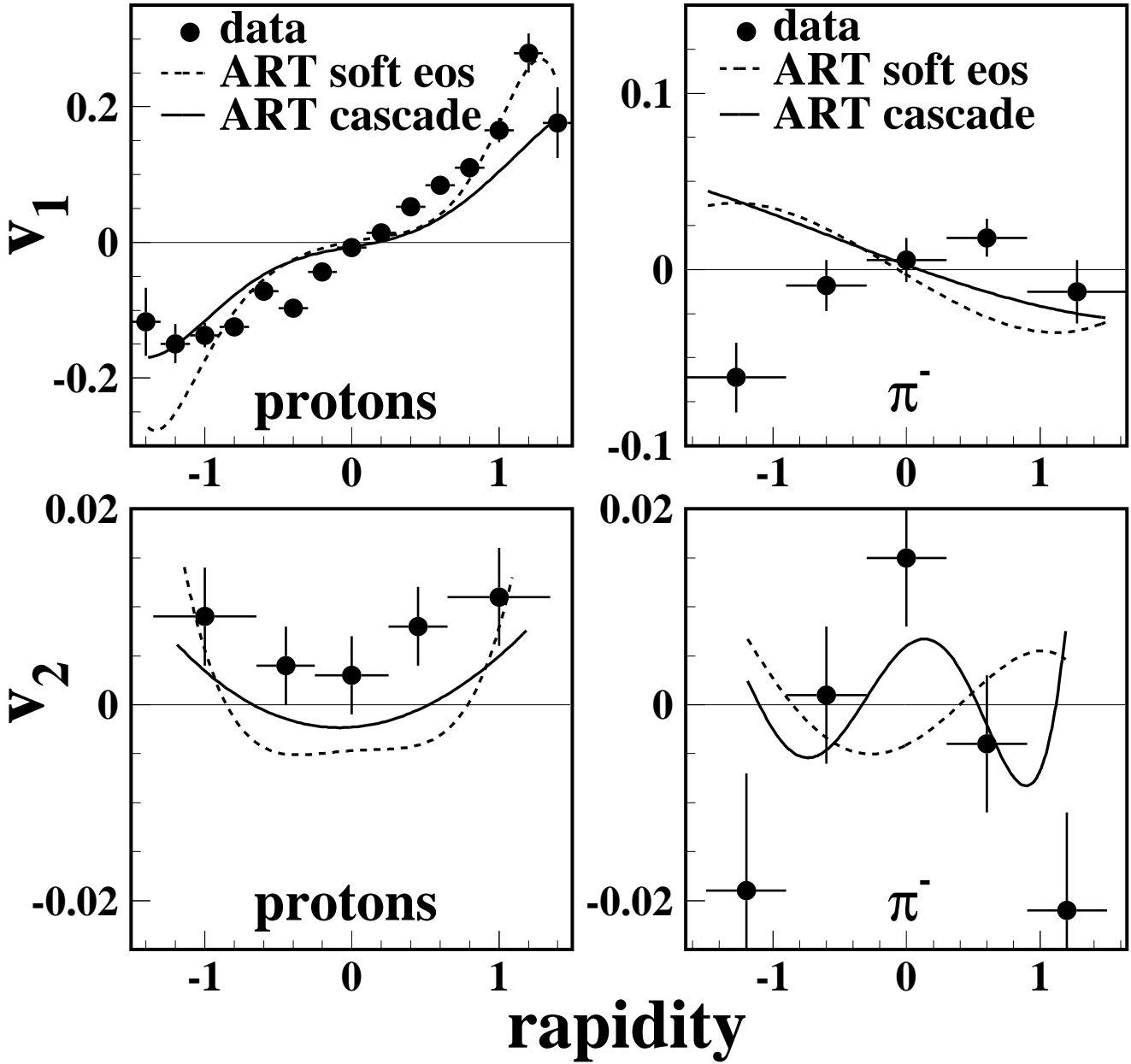
\end{document}